\documentstyle[12pt,amssymb]{article}

\newcommand{\beq}{\begin{equation}}
\newcommand{\eeq}{\end{equation}}
\newcommand{\bea}{\begin{array}}
\newcommand{\eea}{\end{array}}
\newcommand{\bey}{\begin{eqnarray}} 
\newcommand{\pslash}{\not{\hbox{\kern-2.3pt $p$}}}
\newcommand{\pdslash}{\not{\hbox{\kern-2pt $\partial$}}}
\newcommand{\bfx}{{\bf x}}
\newcommand{\bfy}{{\bf y}}
\newcommand{\hatpd}{{\hat{\pd}}}
\newcommand{\hatC}{{\hat C}}
\newcommand{\hatpsi}{{\hat{\psi}}}
\newcommand{\hatp}{{\hat{p}}}

\newcommand{\0}{{\bf 0}_{4\times 4}}

\newcommand{\psibar}{{\overline{\psi}}}

\newcommand{\psitilde}{{\tilde{\psi}}}
\newcommand{\psitildebar}{{\tilde{{\overline{\psi}}}}}

\newcommand{\bp}{{\bf p}}
\newcommand{\ri}{{\rm i}}

\newcommand{\pd}{\partial}

\newcommand{\bfn}{{\mathbf{\nabla}}}
\newcommand{\eey}{\end{eqnarray}}

\def\JMP#1#2#3#4{#2 #1 {\em J. Math. Phys.} {\bf #3} #4}
\def\PRSL#1#2#3#4{#2 #1 {\em Proc. Roy. Soc. (London)} {\bf #3} #4}
\def\HPA#1#2#3#4{#2 #1 {\em Helv. Phys. Acta} {\bf #3} #4}

\def\PRD#1#2#3#4{#2 #1 {\em Phys. Rev.} {\bf D#3} #4}
\def\KDVSMFM#1#2#3#4{#2 #1 {\em Kgl. Dansk. Vidensk. Selsk. Mat. Fys. Medd.}
 {\bf #3} #4}

\def\FP#1#2#3#4{#2 #1 {\em Fortschr. Phys.} {\bf #3} #4}
\def\FoP#1#2#3#4{#2 #1 {\em Found. Phys.} {\bf #3} #4}
\def\AP#1#2#3#4{#2 #1 {\em Ann. Phys.} {\bf #3} #4}
\def\APP#1#2#3#4{#2 #1 {\em Acta Phys. Pol.} {\bf #3} #4}

\def\PR#1#2#3#4{#2 #1 {\em Phys. Rev.} {\bf #3} #4}

\begin{document}

\begin{titlepage}
\vskip 2cm
\begin{center}
{\Large\bf CPT theorem in a $(5+1)$ Galilean space-time 
\footnote{{\tt masanori@gifu-u.ac.jp,}
 {\tt montigny@phys.ualberta.ca,}  {\tt khanna@phys.ualberta.ca} }} 
\vskip 3cm
{\bf 
M. Kobayashi$^{a,b}$, M. de Montigny$^{a,c}$, F.C. Khanna$^{a,d}$ \\} 
\vskip 5pt
{\sl $^a$Theoretical Physics Institute, University of Alberta\\
 Edmonton, Alberta, Canada T6G 2J1\\}
\vskip 2pt
{\sl $^b$Department of Physics, Gifu University\\
  Gifu, Japan 501-1193\\}
\vskip 2pt
{\sl $^c$Campus Saint-Jean, University of Alberta \\
 Edmonton, Alberta, Canada T6C 4G9\\}
\vskip 2pt
{\sl $^d$TRIUMF, 4004, Wesbrook Mall\\
Vancouver, British Columbia, Canada  V6T 2A3\\}
\vskip 2pt

\end{center}
\vskip .5cm
\rm

\begin{abstract}
We extend the 5-dimensional Galilean space-time to a $(5+1)$ Galilean
 space-time in order to define a parity transformation in a covariant
 manner.  This allows us to discuss the discrete symmetries in the
 Galilean space-time, which is embedded in the $(5+1)$ Minkowski
 space-time. We discuss the Dirac-type field, for which we give the
 $8\times 8$ gamma matrices explicitly.  
 We demonstrate that the CPT theorem holds in the
 $(5+1)$ Galilean space-time.  
\end{abstract}



\end{titlepage}

\setcounter{footnote}{0} \setcounter{page}{1} \setcounter{section}{0} %
\setcounter{subsection}{0} \setcounter{subsubsection}{0}




\section{Introduction}
Galilean covariance is feasible in an enlarged $(4+1)$-dimensional manifold, which is embedded
 in the $(4+1)$ Minkowski space-time \cite{omote,others}.
 A 4-dimensional realization of the Clifford algebra in
 the $(4+1)$ Minkowski space-time requires $\gamma^5$ as a fourth ``spatial'' 
 element of $\gamma_\mu$s. In
 order to define a $\gamma^5$-like matrix, it is necessary to extend the theories to a
 $(5+1)$-dimensional manifold \cite{r1}. In this paper, we shall show that it is possible to define,
 in a covariant manner,
 the parity matrix, as well as the charge-conjugation and time-reversal matrices
 in a $(5+1)$ Galilean space-time. Hence we can discuss a CPT theorem based on the transformation
 properties of the Dirac-type field and state. 

In the context of axiomatic field theory, the CPT theorem was established by Wightman \cite{r9}
 and his associates \cite{r10}. In relativistic quantum field theory, Schwinger proved the spin-statistics
 connection by symmetrizing (anti-symmetrizing) the kinematical term of the Lagrangian and, hence,
 the commutation (anticommutation) relations \cite{r7,S2}. Recently, Puccini and Vucetich axiomatized
 Schwinger's Lagrangian formulation and proved the CPT theorem by assuming a form of dynamical
 Lagrangian \cite{r8}. Weinberg adopted the causality requirement, originally proposed by Pauli
 \cite{r6}, to prove the connection between spin and statistics
 as well as the CPT theorem \cite{r9}. 

In Ref. \cite{r1}, we have developed an 8-dimensional
 realization of the Clifford algebra
 in the $(4+1)$ Galilean space-time, in order to define the discrete
 symmetries, in particular, the parity transformation.  This was 
 accomplished by using the dimensional reduction from the $(5+1)$
 Minkowski space-time to the $(4+1)$ Minkowski space-time, which,
 when expressed in terms of light-cone coordinates, corresponds
 to the $(4+1)$ Galilean space-time.

In order to avoid using projective representations of the Galilei
 group, which arise from the existence of a central charge, the
 $(4+1)$ Galilean space-time was
 introduced as the realization of a central
 extension of the group. The usual $(3+1)$ space-time
 is embedded into the $(4+1)$ Galilean space-time.  In 
 odd-dimensional Minkowski space-times, in which the number of spatial 
 coordinates is even, the reflection of spatial manifold has a 
 determinant equal to one, so that it is continuously connected
 to the identity and can be obtained as a rotation.  

Parity refers to a reversal of orientation of the spatial coordinates.  
 Thus we define the parity transformation by the mapping:
\[
P':\quad x^\mu\rightarrow x'^\mu=(-\bfx, x^4, x^5).
\]
However, the existence of this discrete transformation
 entails the loss of manifest covariance, because the space reflection
 in the $(4+1)$ Minkowski space-time corresponds to 
\[
P:\quad x^\mu\rightarrow x'^\mu=(-\bfx, x^5, x^4). 
\]
A way to preserve both manifest covariance and the discrete
 parity operation is to extend the $(4+1)$ Galilean
 space-time to a $(5+1)$ Galilean space-time. The latter space-time
 corresponds
 to a $(5+1)$ Minkowski space-time defined with light-cone
 coordinates.  Motivated by this fact we develop, in this paper,
 a 6-dimensional Galilean theory in a covariant manner
 and prove the CPT theorem.

For the $(5+1)$ Galilean space-time, we use  light-cone coordinates
 $x^\mu$ ($\mu=1,\dots,6$), with the metric tensor
\[
\eta_{\mu\nu}=\left(
\begin{array}{cc}
1_{4\times 4} & 0_{4\times 2}\\
0_{2\times 4} &
\begin{array}{cc}
 0 & -1\\
 -1 & 0\end{array}
\end{array}
\right).
\]
Then the coordinate system $y^\mu$ ($\mu=1,\dots,5, 0$), defined by
\begin{equation}
\bfy=\bfx,\quad
y^4=x^4,\quad
y^5=\frac 1{\sqrt{2}}(x^5-x^6),\quad
y^0=\frac 1{\sqrt{2}}(x^5+x^6), 
\label{e4}
\end{equation}
admits the diagonal metric 
\[
g_{\mu\nu}={\rm{diag}}(1, 1, 1, 1, 1, -1).
\]
This correspondence between the $(5+1)$ Galilean space-time and
 the $(5+1)$ Minkowski space-time allows
 us to describe non-relativistic theories in a 
 Lorentz-like covariant manner \cite{omote,others}.

Let $\Gamma^\mu$ and $\gamma^\mu$ denote the $8\times 8$
 gamma matrices in the $(5+1)$ Galilean (light-cone coordinates $x$s) and
 Minkowski space-times (coordinates $y$s), respectively.  The gamma matrices
 transform as contravariant vectors in each space-time. Therefore,
 we have 
\[
\Gamma^k=\gamma^k=
\left(
\begin{array}{cc}
0_{4\times 4} & 
\begin{array}{cc}
 0 & \sigma_k\\
 -\sigma_k & 0\end{array}
\\
\begin{array}{cc}
 0 & \sigma_k\\
 -\sigma_k & 0\end{array}
& 0_{4\times 4}
\end{array}
\right),\qquad k=1, 2, 3,
\]

\[
\Gamma^4=\gamma^4=
\ri \left(
\begin{array}{cc}
0_{4\times 4} &
\begin{array}{cc}
 0 & I\\
 I & 0\end{array}
\\
\begin{array}{cc}
 0 & -I\\
 -I & 0\end{array}
& 0_{4\times 4}
\end{array}
\right),
\]

\[
\Gamma^5=\frac 1{\sqrt{2}}(\gamma^5+\gamma^0)=
-\sqrt{2}\; \ri\;
 \left(
\begin{array}{cc}
0_{4\times 4} &
\begin{array}{cc}
 0 & 0\\
 0 & I\end{array}
\\
\begin{array}{cc}
 I & 0\\
 0 & 0\end{array}
& 0_{4\times 4}
\end{array}
\right),
\]

\[
\Gamma^6=\frac 1{\sqrt{2}}(-\gamma^5+\gamma^0)=
-\sqrt{2}\; \ri\;
 \left(
\begin{array}{cc}
0_{4\times 4} &
\begin{array}{cc}
 I & 0\\
 0 & 0\end{array}
\\
\begin{array}{cc}
 0 & 0\\
 0 & I\end{array}
& 0_{4\times 4}
\end{array}
\right),
\]

\begin{equation}
\zeta=\frac 1{\sqrt{2}}\ri (\Gamma^5+\Gamma^6)= \ri\gamma^0=
 \left(
\begin{array}{cc}
0_{4\times 4} &
\begin{array}{cc}
 I & 0\\
 0 & I\end{array}
\\
\begin{array}{cc}
 I & 0\\
 0 & I\end{array}
& 0_{4\times 4}
\end{array}
\right),
\label{zeta}
\end{equation}

\[
\Gamma^7=\gamma^7=\gamma^1\gamma^2\gamma^3\gamma^4\gamma^5\gamma^0=
\left(
\begin{array}{cc}
\begin{array}{cc}
 I & 0\\
 0 & I\end{array}
& 0_{4\times 4} \\
0_{4\times 4} &
\begin{array}{cc}
- I & 0\\
 0 & -I\end{array}
\end{array}
\right),
\]
where
\[
I=\left (\begin{array}{cc}1&0\\0&1\end{array}\right ),\ \sigma_1 =
\left (\begin{array}{cc}0&1\\1&0\end{array}\right ),\
  \sigma_2 =\left (\begin{array}{cc}0&-\ri\\\ri&0\end{array}\right ),
\ \sigma_3 =\left (\begin{array}{cc}1&0\\0&-1\end{array}\right ).
\]
Note that we have inverted $\gamma^4$ and $\gamma^5$ with respect 
 to Ref. \cite{r1}.

The canonical conjugate variable of the extended coordinates in
 the $(5+1)$ Galilean space-time provides a transparent interpretation
 of the additional parameter $s$. 
 Indeed, the 6-momentum,
\[
\begin{array}{rcl}
p_\mu=-\ri\pd_\mu&=&(-\ri\bfn, -\ri\pd_4, -\ri\pd_t, -\ri\pd_s),\\
 &=&(\bp, p_4, -E, -m),
\end{array}
\]
such that $p^5=-p_6=m$ and $p^6=-p_5=-E$, shows that the coordinate $s$ is
 conjugate to the mass $m$ in the same way that $\bfx$ is conjugate
 to the momentum $\bp$.

\section{Discrete symmetries}

The parity and time-reversal transformations in the $(5+1)$
 Minkowski space-time are defined respectively by
\[
P: y^\mu \rightarrow y'^\mu=(-\bfy, -y^4, -y^5, y^0)=-g_{\mu\nu}y^\nu,
\]
\[
T: y^\mu \rightarrow y'^\mu=(\bfy, y^4, y^5, -y^0)=g_{\mu\nu}y^\nu,
\]
which correspond to  
\begin{equation}
P: x^\mu \rightarrow x'^\mu=(-\bfx, -x^4, x^6, x^5)=-\eta_{\mu\nu}x^\nu,
\label{e15}
\end{equation}
\begin{equation}
T: x^\mu \rightarrow x'^\mu=(\bfx, x^4, -x^6, -x^5)=\eta_{\mu\nu}x^\nu,
\label{e16}
\end{equation}
in the $(5+1)$ Galilean space-time.  The transformation matrices
 for parity and time reversal, denoted by $R_A$ ($A=P, T$), are
 defined by imposing that the Dirac (and Dirac-like) equation be invariant under
 the transformations $A=P, T$.  The Dirac equation and the Dirac-like
 equation are equivalent to each other under the conditions
 (\ref{e4}), that is,
\[
-(\ri\gamma\cdot\hatpd+\kappa_m)\hatpsi(y)=
 -(\ri \Gamma\cdot\pd+\kappa_m)\psi(x)=0,
\]
where
\[
\hatp_\mu=-\ri\hatpd_\mu=-\ri\frac{\pd}{\pd y^\mu},
\]
and
\[
\hatp_\mu\hatp^\mu=p_\mu p^\mu=-\kappa_m^2,
\]
with 
\[
\kappa_m=\sqrt{2}\; m.
\]

This leads to the fact that if transformation properties under $A=P, T$ are obtained
 in one 6-dimensional space-time, then the same transformation properties hold in the
 other 6-dimensional space-time, in which the metric tensors are interchanged:
 $g_{\mu\nu}\leftrightarrow\eta_{\mu\nu}$.  Henceforth, we discuss the transformation
 properties in the $(5+1)$ Galilean space-time.

The transformation matrices $R_A$ (where $A=P, T$) are obtained by imposing 
\[
-R_A^{-1}\; (\Gamma\cdot\pd'+\kappa_m)\; R_A =
 -(\Gamma\cdot\pd+\kappa_m),\quad\qquad A=P,T,
\]
where
\begin{equation}
\begin{array}{l}
R_P^{-1}=\zeta^{-1}R_P^\dagger\zeta,\\
R_T^{-1}=-\zeta^{-1}R_T^\dagger\zeta,
\end{array}\label{raprat}
\end{equation}
and $\zeta$, defined as in Eq. (\ref{zeta}), satisfies
 $\zeta=\zeta^{-1}$.
The notation $\pd'$ is defined by
\[
\pd'_\mu=\frac\pd{\pd x'^\mu},
\]
where $x'^\mu$ is given by Eq. (\ref{e15}) or (\ref{e16}),
 depending on whether $A=P$ or $T$.
 The negative sign in the second expression of Eq. (\ref{raprat})
 arises from the fact that Dirac-type (Dirac) field should obey anti-commutation
 relations.  It is important to mention that for a scalar-type (scalar) field and a vector-type (vector) field, this sign will be positive.

Explicitly, we find that
\[
\begin{array}{l}
R_P=\gamma^0=\frac 1{\sqrt{2}}(\Gamma^5+\Gamma^6),\\
R_T=\gamma^1\gamma^2\gamma^3\gamma^4\gamma^5=\gamma^0\gamma^7=
\frac 1{\sqrt{2}}(\Gamma^5+\Gamma^6)\Gamma^7.
\end{array}
\] 

The charge-conjugation matrix $C$ is defined by
\begin{equation}
(\Gamma^\mu)^T=-C^{-1}\Gamma^\mu C=
 \hatC^{-1}(\Gamma^\mu)^\dagger\hatC,
\label{e29}
\end{equation}
with 
\[
\hatC^\dagger=\hatC^{-1}=-\hatC^\ast,\qquad\quad
\hatC^T=-\hatC,
\]
where
\begin{equation}
\begin{array}{l}
\hatC=\gamma^0 C=\frac 1{\sqrt{2}} (\Gamma^5+\Gamma^6)C,\\
(\Gamma^\mu)^\dagger=\frac 12 (\Gamma^5+\Gamma^6)\Gamma^\mu (\Gamma^5+\Gamma^6).
\end{array}
\label{e32}
\end{equation}
The charge-conjugation matrix $\hatC$ satisfies
\[
[-\zeta\; (\Gamma\cdot\pd+\kappa_m)]^T
=\{-\zeta\; [\Gamma\cdot(-\pd)+\kappa_m]\}^\ast
=\hatC^{-1}\zeta\; [\Gamma\cdot(-\pd)+\kappa_m]\; \hatC,
\]
which reduces to
\[
-C [\Gamma\cdot(-\pd)+\kappa_m]^T\; C^{-1}=-(\Gamma\cdot\pd+\kappa_m).
\]

Galilean theory is manifestly covariant under Galilean transformations. Therefore
 equations remain valid in all frames of reference.  This statement
 does not make sense, however, unless transformations are adequately
 assigned to the fields. Thus, hereafter we define such field
 transformations.

Now let us define the $C$, $T$ and $P$
 transformation properties of Dirac-type fields.  
 For charge conjugation $C$, we have:
\begin{equation}
\begin{array}{l}
\psi'(x')=\xi_C\psi_C(x)=\xi_C\hatC\psi^\ast(x),\\
\psibar'(x')=\psi'^\dagger(x')\zeta=\xi^\ast_C\psibar_C(x)=
\xi_C^\ast\psi^T(x)\hatC^\dagger\zeta,
\end{array}
\label{e36}
\end{equation}
with
\[
\hatC:\Psi\rightarrow \Psi'=\Psi,
\]
where the star $\ast$ denotes the complex conjugation of a $c$-number
 (including gamma matrices) and the 
 Hermitian adjoint of an operator. For time-reversal $T$, we define
\begin{equation}
\begin{array}{l}
\psi'(x')=\xi_TR_T\psitilde_C(x),\\
\psibar'(x')=\xi^\ast_T\psitilde_C^T(x)
\zeta^{-1} R_T^\dagger\zeta,
\end{array}
\label{e37}
\end{equation}
with
\[
T:\Psi\rightarrow \Psi'=\Psi^\ast,
\]
where the tilde $\ \tilde{}\ $ denotes the transposition of an
 operator in $\psi_C(x)$. For a parity transformation $P$, we have 
\begin{equation}
\begin{array}{l}
\psi'(x')=\xi_PR_P\psi(x),\\
\psibar'(x')=\xi^\ast_P\psibar(x)
\zeta^{-1} R_P^\dagger\zeta,
\end{array}
\label{e38}
\end{equation}
with
\[
P:\Psi\rightarrow \Psi'=\Psi.
\]
Here $\Psi$ stands for the state vector and the complex $c$-number
 $\xi_A$ (where $A=C, P, T$) can be normalized to unity without loss
 of generality:
\[
|\xi_A|^2=\xi_A\xi^\ast_A=1.
\]

Now that we have assigned the transformation properties of the 
 discrete symmetries to the fields, the transformations of the
 bilinear forms under $\hatC$, $P$ and $T$ will be obtained
 below.


\subsection{Charge conjugation}

Recalling the definitions of $C$ and $\hatC$ in Eqs.
 (\ref{e29}) to (\ref{e32}), we find 
\[
\psibar'(x')\Gamma^A\psi'(x')=-\epsilon^A\psibar(x)\Gamma^A\psi(x),
\]
with
\[
\epsilon^A=\left\{
\begin{array}{l}
+1,\qquad {\rm{for}}\ \Gamma^A=1, \Gamma^7\Sigma^{\mu\nu}, 
 \Sigma^{\lambda\mu\nu},\\
-1,\qquad {\rm{for}}\ \Gamma^A=\Gamma^7, \Gamma^\mu, \ri\Gamma^7\Gamma^\mu,
 \Sigma^{\mu\nu},\end{array}\right.
\]
where
\[
\begin{array}{l}
\Sigma^{\mu\nu}=\frac 1{2\ri}(\Gamma^\mu\Gamma^\nu-\Gamma^\nu\Gamma^\mu),\\
\Sigma^{\lambda\mu\nu}=\frac 13(\Gamma^\lambda\Sigma^{\mu\nu} +
 \Gamma^\mu\Sigma^{\nu\lambda}+ \Gamma^\nu\Sigma^{\lambda\mu}).
\end{array}
\]
Here we have utilized the relationship:
\[
\zeta^{-1}\; (\hatC^\dagger\zeta\Gamma^A\hatC)^T=
 -(C^{-1}\Gamma^AC)^T=-\epsilon^A\Gamma^A.
\]

The commutation relation becomes
\begin{equation}
\{\psi_{C\alpha}(x_1), \psibar_C^{\ \ \beta}(x_2)\}=
\ri\; d_\alpha^{\ \ \beta}(\pd_1)\Delta(x_1-x_2),
\label{e47}
\end{equation}
where
\[
d(\pd)=-(\Gamma\cdot\pd-\kappa_m).
\]
The commutation relation in Eq. (\ref{e47}) is invariant under
 the charge-conjugation transformation of Eq. (\ref{e36}). 
 Therefore, there exists a unitary transformation which
 is called the `charge-parity operator', such that
\[
\begin{array}{l}
G_C^{-1}\psi(x)G_C=\xi_C\hatC\psi^\ast(x),\\
G_C^{-1}\psibar(x)G_C=\xi_C^\ast\psi^T(x)\hatC^\dagger\zeta,
\end{array}
\]
with
\[
G_C|0\rangle = |0\rangle .
\]

\subsection{Time reversal}

The bilinear forms transform under the time-reversal transformation, Eq. (\ref{e37}), as
\[
\begin{array}{rcl}
\psibar'(x')\Gamma^A\psi'(x')&=&
 -\psitildebar(x)\zeta^{-1}\;
 (\hatC^\dagger\zeta R_T^{-1}\Gamma^AR_T\hatC)^T\; \psitilde(x),\\
 &=&\epsilon^T\epsilon^A\; \psitildebar(x)\Gamma^A\psitilde(x),
\end{array}
\]
with
\begin{equation}
\epsilon^T=\left\{
\begin{array}{cl}
1, & {\rm{for}}\ \Gamma^A=1,\\
-1, & {\rm{for}}\ \Gamma^A=\Gamma^7,\\
\eta_{\mu\rho}, & {\rm{for}}\ \Gamma^A=\Gamma^\mu,\\
-\eta_{\mu\rho}, & {\rm{for}}\ \Gamma^A=\ri\Gamma^7\Gamma^\mu,\\ 
\eta_{\mu\rho}\eta_{\nu\sigma}, & {\rm{for}}\ \Gamma^A=\Sigma^{\mu\nu},\\
-\eta_{\mu\rho}\eta_{\nu\sigma}, & {\rm{for}}\ \Gamma^A=\Gamma^7\Sigma^{\mu\nu},\\
\eta_{\lambda\rho}\eta_{\mu\sigma}\eta_{\nu\tau},
 & {\rm{for}}\ \Gamma^A=\Sigma^{\lambda\mu\nu},\end{array}\right.
\label{e52}
\end{equation}
which reads, for example,
\[
\psibar'(x')\Gamma^\mu\psi'(x')=\eta_{\mu\rho} (-1)
\psitildebar(x)\Gamma^\rho\psitilde(x).
\]
It should be remarked that a physical quantity $A$, 
 expressed in functional form, is invariant under the
 time-reversal transformation in the sense:
\[
A[\psi'(x'),\pd'_\mu\psi'(x'),\dots]=
{\tilde{A}}[\psi(x),\pd_\mu\psi(x),\dots].
\]

The commutation relation is obtained from Eqs. (\ref{e37}) and
 (\ref{e47}):
\[
\{\psi'_\alpha(x'_1), \psibar'^\beta(x'_2)\}
=\ri\; d_\alpha^{\ \ \beta}(\pd'_1)\;
 \Delta(x'_1-x'_2).
\]
The invariance of the commutation relations implies that a
 unitary transformation exists:
\[
\begin{array}{l}
G_T^{-1}\psi(x)G_T=\xi_TR_T\psitilde_C(x'),\\
G_T^{-1}\psibar(x)G_T=\xi_T^\ast\psitildebar_C(x')\;
 (\zeta^{-1}R_T^\dagger\zeta),
\end{array}
\]
with 
\[
G_T|0\rangle = |0\rangle,
\]
and $x'$ given by Eq.\,(\ref{e16}).

If we define 
\[
\Psi^T=G_T\Psi^\ast=G_T\Psi',
\]
then
\[
\langle\Psi'_2,G_T^{-1}\psi(x)G_T\Psi'_1\rangle=
\langle\Psi^T_2,\psi(x)\Psi^T_1\rangle=
\langle\Psi_2^\ast,\psitilde(x')\Psi_1^\ast\rangle =
 \langle\Psi_1,\psi(x')\Psi_2\rangle .
\]


\subsection{Space reflection}

The bilinear forms transform under the space reflection
 transformation, Eq. (\ref{e38}), as
\[
\begin{array}{rcl}
\psibar'(x')\Gamma^A\psi'(x')&=&\psibar(x)R_P^{-1}\Gamma^AR_P\psi(x),\\[1ex]
 &=&\epsilon^P\; \psibar(x)\Gamma^A\psi(x),
\end{array}
\]
where $\epsilon^P$ is obtained by the replacement 
\[
\eta_{\mu\nu}\rightarrow -\eta_{\mu\nu},
\]
in $\epsilon^T$ of Eq.\,(\ref{e52}). This implies the important fact that
 vectors and rank-3 tensors change signs, whereas the signs
 of other quantities remain unchanged under successive discrete 
 transformations $\hatC$, $T$ and $P$.  Thus, the $CPT$ theorem
 holds in the $(5+1)$ Galilean space-time.

The commutation relation is expressed in the form 
\[
\{\psi'_\alpha(x'_1), \psibar'^\beta(x'_2)\}=\ri\;
 d_\alpha^{\ \ \beta}(\pd'_1)\Delta(x'_1-x'_2),
\]
which is invariant under the parity transformation, Eq. (\ref{e38}). 
This means that a unitary transformation $G_P$ (the parity operator) exists, such that
\[
\begin{array}{l}
G_P^{-1}\psi(x)G_P=\xi_PR_P\psi(x'),\\
G_P^{-1}\psibar(x)G_P=\xi_P^\ast\psibar(x')R_P^{-1},
\end{array}
\]
with
\[
G_P|0\rangle = |0\rangle,
\]
and $x'$ given in Eq. (\ref{e15}).  
In deriving the relations developed in this section, the vacuum subtraction should
 be understood.

As stated earlier, we can develop the same procedure in the
 $(5+1)$ Minkowski space-time and the same results are obtained
 by replacing $\eta_{\mu\nu}$ with $g_{\mu\nu}$. The Dirac gamma
 matrices are given by using the same charge-conjugation matrix as 
 in Eq. (\ref{e29}):
\[
\gamma^A=\epsilon^A\; (C^{-1}\gamma^AC)^T,
\]
with  
\[
\epsilon^A=\left\{
\begin{array}{l}
+1,\qquad {\rm{for}}\ \gamma^A=1, \gamma^7\sigma^{\mu\nu},
 \sigma^{\lambda\mu\nu},\\
-1,\qquad {\rm{for}}\ \gamma^A=\gamma^7, \gamma^\mu, \ri\gamma^7\gamma^\mu,
 \sigma^{\mu\nu},\end{array}\right.
\]
where
\[
\begin{array}{l}
\sigma^{\mu\nu}=\frac 1{2\ri}(\gamma^\mu\gamma^\nu-\gamma^\nu\gamma^\mu),\\
\sigma^{\lambda\mu\nu}=\frac 13(\gamma^\lambda\sigma^{\mu\nu} +
 \gamma^\mu\sigma^{\nu\lambda}+ \gamma^\nu\sigma^{\lambda\mu}).
\end{array}
\]


\section{Concluding remarks}

In relativistic quantum field theory, the CPT theorem was developed in
 connection with spin and statistics \cite{r3,r4}.   L\"uders gave a
 proof of the CPT theorem by using the transformation properties
 of each kind of field and hence a connection between spin and
 statistics was postulated \cite{r4}.  In fact, he made the proof
 for the fields with spin 0, 1/2 and 1, and concluded that the proof
  can be extended to fields with arbitrary spin, which are constructed
 in terms of spin $1/2$ and 1 fields. 

The conclusion that the CPT theorem holds in the $(5+1)$ Galilean space-time
 is easily extended to the scalar- and vector-type fields, since the proof given here
 is based on the transformation properties of the field. Thus the CPT 
 theorem in $(5+1)$ Galilean space-time holds, in the
 sense of L\"uders. Furthermore it may be stated, following Schwinger \cite{r7},
 that ``connection between spin and statistics of particles is
 implicit in the requirement of invariance under coordinate transformations''.


\section*{Acknowledgment}

We acknowledge partial support by the Natural Sciences and Engineering
 Research  Council of Canada. This manuscript was completed while M.K. was
 visiting the Theoretical Physics Institute at the University of Alberta.


\end{document}